\documentstyle[preprint,aps]{revtex}
\begin{document}
\draft 
\preprint{IFUSP-P/1257, hep-th/9702065}

\title{Duality Symmetry in the Schwarz-Sen Model\cite{byline}} 

\author{H. O. Girotti$^a$, M. Gomes$^b$, V. O. Rivelles$^b$ and A. J.
  da Silva$^b$} \address{$^a$ Instituto de F\'{\i}sica, Universidade
  Federal do Rio Grande do Sul, CP15051, 91501-970 Porto Alegre, RS, Brazil.\\
  $^b$ Instituto de F\'\i sica, Universidade de S\~ao Paulo, CP66318,
  05315--970, S\~ao Paulo, SP, Brazil.}

\maketitle

\begin{abstract}
  The continuous extension of the discrete duality symmetry of the
  Schwarz-Sen model is studied. The corresponding infinitesimal
  generator $Q$ turns out to be local, gauge invariant and metric
  independent. Furthermore, $Q$ commutes with all the conformal group
  generators. We also show that $Q$ is equivalent to the non---local
  duality transformation generator found in the Hamiltonian formulation
  of Maxwell theory. We next consider the Batalin--Fradkin--Vilkovisky
  formalism for the Maxwell theory and demonstrate that requiring a local
  duality transformation lead us to the Schwarz--Sen formulation. The
  partition functions are shown to be the same which implies  the quantum
  equivalence of the two approaches.
\end{abstract}

\narrowtext
\newpage

It is nowadays accepted that the known string theories are different
perturbative versions of an underlying M-theory \cite{1}. This
idea was originated by the several duality symmetries present in
the string theories. It is therefore important to study in detail
duality symmetries in field theory and understand its implications. In
this respect the old electric-magnetic duality present in Maxwell's
equations has again  been the subject of intensive study recently 
\cite{2,3,4,5,6,7}.  These studies show
that there is a conflict between electric-magnetic duality symmetry
and manifest Lorentz covariance when we attempt to implement duality
at the action level. If manifest Lorentz covariance is maintained then
the action is either non-polynomial \cite{2} or requires an infinite
set of fields \cite{3}. If we give up manifest Lorentz covariance then
duality symmetry can be implemented in the Hamiltonian formulation of
Maxwell theory in a non-local way \cite{4}. However a recent proposal
made by Schwarz and Sen implements duality in a local way at the
expenses of introducing one more potential \cite{5}. Although
Schwarz--Sen's formulation is not manifestly covariant it is
Poincar\'e covariant both at the classical \cite{5} and quantum
\cite{7} levels.  In this paper we shall investigate some consequences
of such proposal.  In particular, we will construct the generator of
duality transformations and discuss its meaning and relation with the
corresponding non--local generator found in the first order Hamiltonian
formulation described in \cite{4}. We will also verify that Schwarz--Sen's
proposal can be understood as a formulation where the non-local duality
transformation of \cite{4} is turned into a local one. This is shown in the
Batalin--Fradkin--Vilkovisky path integral formalism. As a consequence, this
also implies in the quantum equivalence of the two approaches.
Although we consider only the free field case, our work exemplifies the use
of a method that may be helpful in more general situations.

The Schwarz-Sen action \cite{5} involves two gauge potentials $A^{\mu a}$
($1\leq a \leq 2$ and $0\leq \mu\leq 3$) and is given by 

\begin{equation}\label{1}
S = -\frac{1}{2} \int d^4 x ( B^{a,i} \epsilon^{ab} E^{b,i} + B^{a,i} B^{a,i}),
\end{equation}

\noindent
where 
\begin{eqnarray}\label{2}
E^{a,i}& = &-F^{a,0i}=- (\partial^0 A^{a,i} - \partial^i A^{a,0}),\\
B^{a,i}& = &-\frac{1}{2} \epsilon^{ijk} F^{a}_{jk}=-\epsilon^{ijk}
\partial_j A^{a}_{k}
\end{eqnarray}

\noindent
and $\epsilon$ is the Levi-Civita symbol ($\epsilon^{12}=1, \epsilon^{123}=1$)
and $1\leq i,j,k\leq 3$.  This action is separately invariant under
the local gauge transformations,

\begin{eqnarray}\label{3}
A^{a,0} &\rightarrow & A^{a,0} + \psi^a, \\
A^{a,i} &\rightarrow &A^{a,i} - \partial^i\Lambda^a
\end{eqnarray}

\noindent
and the global SO(2) rotations,
\begin{equation}\label{4}
A^{\prime\mu a} =  A^{\mu a}\cos \theta + \epsilon^{a b} A^{\mu b} \sin \theta,
\end{equation}

\noindent
which reduces to the usual discrete duality transformation for $\theta=\pi/2$.

The Noether's charge associated with this SO(2) symmetry is
\begin{equation}\label{5}
Q=-\frac{1}{2}\int d^3 x \epsilon^{jik} (\partial_j A^{a}_i) A_{k}^{a} = 
\frac{1}{2}\int d^3x B^{a k} A^{a}_{k}.
\end{equation}

\noindent
Notice that $Q$ is a SO(2) invariant Chern-Simons term. Hence, up to
surface terms, it is gauge invariant. It is also metric independent
and so its algebraic form also holds for curved spaces.

Using the Coulomb gauge equal time commutators  \cite{7},
\begin{equation}\label{6}
[A^{a,i}(\vec x), A^{b,j}(\vec y)]= -i \epsilon_{ab} \epsilon^{ijk}\frac 
{\partial^{x}_{k}}{\nabla^2} \delta (\vec x - \vec y),
\end{equation}

\noindent
it is straightforward to verify that $Q$ indeed generates infinitesimal
$SO(2)$ rotations,
\begin{equation}\label{7}
[Q, A^{b}_{j}(y)]= - i \epsilon^{ba} A^{a}_{j}(y).
\end{equation}

The Fock space of states is constructed through the action of creation
and annihilation operator $a^{\dag}_{\lambda}$ and $a_\lambda$ introduced
via the Fourier decomposition of $A^{a,i}$,
\begin{equation}\label{8}
A^{a,i}(x)= \frac{1}{(2\pi)^{3/2}} \int \frac{d^3p}{\sqrt {2 |\vec p|}} 
\sum_{\lambda=1}^{2}
({\rm e}^{-ipx} \epsilon^{a i}_{\lambda} a_\lambda(p) +{\rm e}^{ipx} 
\epsilon^{a i}_{\lambda} a^{\dag}_{\lambda}(p)) ,
\end{equation}

\noindent
where $px \equiv |\vec p| x^0- \vec p \cdot \vec x$ and
$\epsilon^{a,i}_{\lambda}(p)$, $\lambda =1,2$ are unit norm
polarization vectors, orthogonal to $\vec p$ and satisfying

\begin{equation}\label{9}
(g_{ij}\delta_{ab} p_0 - \epsilon_{ab} \epsilon_{ilj} p^l)
\epsilon^{bj}_{\lambda}(p)=0.
\end{equation}
\noindent
This means that $(\hat p, \epsilon^{a}_{1}(\vec p), \epsilon^{a}_{2}(\vec p))$,
 $a=1,2$, are two ortonormal basis rotated by $\pi/2$ in the direction of 
$\hat p$. 
The operators $a_\lambda$ and $a^{\dag}_{\lambda}$ satisfy the usual
algebra
\begin{equation}
[a_\lambda (\vec p), a^{\dag}_{\lambda'}(\vec {p'})] = \delta_{\lambda 
\lambda'}
\delta(\vec p - \vec {p'})
\end{equation}
\noindent
In terms of these operators the charge $Q$ can be rewritten as
\begin{equation}\label{10}
Q= i\int d^3k (a^{\dag}_1 a_2 - a^{\dag}_2 a_1)
\end{equation}
and becomes diagonal, 
\begin{equation}\label{11}
Q=\int d^3 k (a^{\dag}_{L} a_L - a^{\dag}_{R} a_R)
\end{equation}
in the base of circularly polarized operators, defined by
\begin{eqnarray}\label{12}
a^{\dag}_{R}& = & \frac{a^{\dag}_{1}+i a^{\dag}_{2}}{\sqrt 2} \\
a^{\dag}_{L}& = & \frac{a^{\dag}_{1}-i a^{\dag}_{2}}{\sqrt 2}.
\end{eqnarray}
>From (\ref{11}) one sees that, in a generic state, $Q$ counts the number
of left minus right polarized photons.

It is easily checked that $Q$ commutes with the components
$\theta^{0\mu}$ of the energy momentum tensor. Hence, it commutes
with all the
generators of the
conformal group as should be expected from an internal symmetry.

Identifying the operators $a_\lambda$ with the corresponding ones in
Maxwell's theory one could work backward and find that the charge $Q$
has the non--local expression,
\begin{equation}\label{13}
Q= \frac{1}{2} \int d^3 x ( -\vec A \cdot \nabla \times \vec A + \vec E \cdot
\nabla^{-2} \nabla \times \vec E),
\end{equation}

\noindent
where $\vec E$  is the electric field and $\vec A$ the vector potential in the
Coulomb gauge. As described in \cite{4}, (\ref{13}) arises from a first
order Hamiltonian formulation of Maxwell action. 

The expression (\ref{13}) for the charge $Q$ can also be  arrived through
formal manipulations using the equations of motion. In fact, using the
gauge freedom (\ref{3}) Schwarz and Sen have shown that
\begin{equation}\label{14}
\vec B^2 = \vec E^1.
\end{equation}

\noindent
Thus, taking the curl of this equation and using the Coulomb gauge condition
one has formally
\begin{equation}\label{15}
\vec A^2 = - \nabla^{-2} \nabla \times \vec E^1.
\end{equation}

\noindent
Equation (\ref{13}) follows from the replacement of (\ref{14}) and (\ref{15})
into (\ref{5}).

Our discussion indicates that Deser and Teitelboim and Schwarz and Sen
implementations of duality are equivalent and both formulations have
the same physical content. We now show how this equivalence can be
understood in the path integral framework.  To see that let us use the
Batalin--Fradkin--Vilkovisky formalism \cite{8} for constrained
systems.  The generating functional for the Maxwell theory is
\begin{equation}\label{15a}
Z= \int {\cal D} A_\mu \,{\cal D}\pi_\mu \,{\cal D}c \,{\cal D} 
\bar {\cal P}\,{\cal D}\bar c\, {\cal D}{\cal P} \, {\rm e}^{iS_{eff}}
\end{equation}
with the effective action given by
\begin{equation}\label{16}
S_{eff}= \int d^4x (\pi_i\dot A^i + \pi_0 \dot A^0 + \dot{\cal P}\bar c 
+\dot c\bar
{\cal P} - {\cal H}_0 - \{\Psi, Q_B\}).
\end{equation}

\noindent
As usual, $\pi_\mu$ is the conjugate momentum of $A_\mu$,
$c$ and $\bar c$ are ghosts and ${\cal P}$ and $\bar {\cal P}$ their
conjugate momenta. At equal times they satisfy
\begin{equation}\label{17}
\{\bar {\cal P}(\vec x), c(\vec y)\} =-\delta(\vec x - \vec y)
\qquad \{{\cal P}(\vec x),  
\bar c(\vec y)\} =-\delta(\vec x - \vec y).
\end{equation}
The BRS charge and  Hamiltonian densities   are given by  
\begin{eqnarray}\label{18}
Q_B &=& \int d^3 x \,\,\,\,\partial_i \pi^i c - i {\cal P} \pi_0\\
{\cal H}_0 &=& -\frac{1}{2}  (\pi^i\pi_i + B^i B_i).
\end{eqnarray}

It is convenient to choose the gauge fixing function $\Psi$ as
\begin{equation}\label{19}
\Psi = i\bar c \chi + \bar {\cal P} A_0 
\end{equation}

\noindent
and fix $\chi$ in such way that the Coulomb condition holds. This can
be achieved if $\chi=\frac{1}{\epsilon} \partial_i A^i$, we
redefine the fields $\pi_0\rightarrow \epsilon \pi_0$,  $\bar c 
\rightarrow \epsilon \bar c$ and let $\epsilon$ go to zero. Notice
that this scaling transformation produces a trivial Jacobian and is
compatible with the BRS transformation,
\begin{eqnarray}
\delta_{BRS} A_i&=&\partial_i c, \quad \delta_{BRS} \pi_i=0, \quad 
\delta_{BRS} A_0= i {\cal P},\quad   \delta_{BRS} \pi_0= 0,\\
\delta_{BRS} c &=& 0,\quad
\delta_{BRS} \bar c = i\pi_0, \quad \delta_{BRS} \bar  {\cal P}= - 
\partial_i \pi^i, \quad \delta_{BRS} {\cal P}=0 
\end{eqnarray}

\noindent
which leaves invariant the generating functional $Z$.  After taking the limit,
$\epsilon \rightarrow 0$ we get,

\begin{equation}\label{20}
S_{eff} = \int d^4 x(\pi_i\dot A^i + \dot c \bar {\cal P} - {\cal H}_0 
+ A_0 \partial_i
\pi^i + \pi_0\partial^i A_i + i \bar {\cal P}{\cal P} - i\bar c \nabla^2 c)
\end{equation}

It is interesting to observe that already at this level there is a duality
transformation of the fields  which leaves $Z$ invariant.
The infinitesimal form of the transformation is given by
\begin{equation}\label{20a}
\delta A_i = \theta \nabla^{-2} \epsilon_{ijk} \partial^j \pi^k \quad 
\qquad \delta \pi_i = \theta \epsilon_{ijk} \partial^j A^k 
\end{equation}

\noindent
the variations of the remaining fields being zero. Notice that the
variation of $A_i$ and $\pi_i$ is the non--local duality
transformation introduced by Deser and Teitelboim.  $\delta$ commutes
with the BRS transformation, as can be checked.

Integrating (\ref{15a}) in $\pi_0$ and $A^0$ produces the usual delta
functions which characterize the Coulomb gauge. Performing also the
trivial integrations in the ghosts $\bar {\cal P}$ and ${\cal P}$ we
finally have
\begin{equation}\label{21}
Z = \int {\cal D} A_i\, {\cal D}\pi_i \,{\cal D}c\, {\cal D}\bar c\, 
{\rm e}^{iS_eff}\, \delta(\partial_i A^i)\, \delta(\partial_i\pi^i)
\end{equation}
where now $S_{eff}$ is the action considered by Deser and Teitelboim,
 
\begin{equation}\label{22}
S_{eff}= \int d^4 x (\pi_i\dot A^i - {\cal H}_0- i\bar c \nabla^2 c).
\end{equation}

\noindent
This action is of course invariant under the Deser--Teitelboim duality
transformation, given by (\ref{20a}) and  $\delta c =\delta \bar c =0$.

We will now show that the Schwarz--Sen formulation can be obtained by
requiring that the duality transformation be local. To this end we
introduce two auxiliary fields $C_i$ and $P_i$ so that
\begin{equation}\label{24}
\delta A_i = \theta C_{iT} \quad \qquad
\delta \pi_i = \theta P_{iT}
\end{equation}
where the index $T$ indicates the transversal part of the field, as
required by the Coulomb gauge condition on $A_i$ and $\pi_i$. These
fields are fixed through the equations
\begin{equation}\label{25}
\nabla^2 C_{iT} =  \epsilon_{ijk}\partial^j \pi^k \quad \qquad
P_{iT} =\epsilon_{ijk} \partial^j A^k
\end{equation}
and the transversality conditions $\partial^i C_i=- \partial^i P_i=0$.

From (\ref{25}) and (\ref{20a}) we have,
\begin{equation}\label{26}
\nabla^2 \delta C_{iT}= \theta \epsilon_{ijk}\epsilon^{klm} 
\partial^j\partial_l A_m=-
\theta \nabla^2 A_i.
\end{equation}
Hence, the transversality condition implies that $\delta C_i =- \theta
A_i$.  Furthermore, proceeding in analogous way one can prove that
$\delta P_i=-\theta \pi_i$.

Comparing these equations with the infinitesimal form of (\ref{4}) for
the Schwarz--Sen approach allow us to identify their fields as
follows,
\begin{equation}\label{27}
A^{(1)}_{i} =  A_i  \quad \qquad A^{(2)}_{i} =  C_i. 
\end{equation}

\noindent
The conjugate momenta to these fields are,
\begin{equation}\label{27a}
\pi^{(1)}_{i} = \frac{1}{2} \pi_i \quad \qquad \pi^{(2)}_{i}= \frac{1}{2} P_i
\end{equation}

To construct the Schwarz--Sen action first notice that
\begin{equation}\label{28}
P^i \dot C_i = -\epsilon^{ijk} \partial_j A_k \epsilon_{ilm} \nabla^{-2}
\partial^l \dot \pi^m= -A^i \dot \pi_i 
\end{equation}
\noindent
and
\begin{equation}\label{29}
\epsilon^{ijk} \nabla^2 \partial_j C_{kT}= -\nabla^2 \pi^i
\end{equation}
from which $\pi_i = - \epsilon_{ijk} \partial^j C^{k}_{T}$.
We can therefore write (\ref{22}) as
\begin{equation}\label{30}
S_{eff}= \int d^4 x (\frac{1}{2} \pi_i\dot A^i+\frac{1}{2}P_i\dot C^i  - 
{\cal H}_0- i\bar c \nabla^2c),
\end{equation}
where now ${\cal H}_0=1/2 ((\nabla \times \vec C)^2 +(\nabla \times
\vec A)^2)$.

Up to the ghost term, equation(\ref{30}) is Schwarz--Sen's action in
Hamiltonian form. Although our discussion already demonstrates the
equivalence of the two approaches, we will prove now that the
generating functional $Z$ given by (\ref{21}) is equal to the
corresponding functional obtained using the Batalin--Fradkin--Vilkovisky 
formalism
for the Schwarz--Sen action. After the introduction of the auxiliary
fields $C_i$ and $P_i$, $Z$ is obviously equal to
\begin{eqnarray}
Z &=& \int {\cal D} A_i\, {\cal D}\pi_i \,{\cal D}C_i\,{\cal D}P_i {\cal D}c\,
 {\cal D}\bar c\,\delta (C_{iT} - \nabla^{-2} \epsilon_{ijk} \partial^j \pi^k)
\delta(P_{iT}-\epsilon_{ijk}\partial^j A^k)\times \nonumber \\
&{\phantom a}& \delta(\partial_i A^i)\, \delta(\partial_i\pi^i)\,
\delta(\partial_i C^i) \delta(\partial_i P^i) {\rm e}^{iS_eff}.\label{31}
\end{eqnarray}
Now, we can write
\begin{equation}\label{32}
\delta (C_{iT} - \nabla^{-2} \epsilon_{ijk} \partial^j \pi^k)= 
{\det}^{-1}(\nabla^{-2} \epsilon_{ijk} \partial^j)\delta (\pi_i+
\epsilon_{ijk}\partial^j C^k).
\end{equation}

\noindent
But, because of the transversality property guaranteed by the delta
functions in (\ref{31}), ${\det}^{-1}(\nabla^{-2} \epsilon_{ijk}
\partial^j)=\det(\epsilon_{ijk} \partial^j)$. So,

\begin{eqnarray}
Z &=& \int {\cal D} A_i\, {\cal D}\pi_i \,{\cal D}C_i\,{\cal D}P_i {\cal D}c\,
 {\cal D}\bar c \delta(\partial_i A^i) \,
 \delta(\partial_i\pi^i)\,\times \nonumber \\
&\phantom a& \det(\epsilon_{ijk}\partial^j) \,\delta (\pi_i+
\epsilon_{ijk}\partial^j C^k)\, \delta(P_{iT}-\epsilon_{ijk}\partial^j A^k)
\,\delta(\partial_i C^i) \delta(\partial_i P^i) {\rm e}^{iS_eff} \label{33}
\end{eqnarray}

As shown in \cite{7} the Schwarz--Sen action exhibits both first
 and second class constraints. They are, respectively,
\begin{equation}\label{34}
\Omega^{a}_0 \equiv \pi^{a}_{0} \approx 0, \qquad \qquad 
\Omega^a \equiv \partial^i \pi_{i}^{a} \approx 0
\end{equation}
and 
\begin{equation}\label{35}
\Omega^{a}_{iT} \equiv \pi^{a}_{iT} +\frac {1}{2} \epsilon_{ab} 
\epsilon_{ijk}\partial^j A^{b, k}_{T}\approx 0
\end{equation}
At equal times, the first class constraints satisfy an Abelian algebra
whereas for the second class constraints we have
\begin{equation}\label{35a}
\{ \Omega_{iT}^{a}(\vec x), \Omega_{jT}^{b}(\vec y) \} = -\epsilon_{ab} 
\epsilon_{ijk} \partial^{k}_{x} \delta (\vec x - \vec y)
\end{equation}
Therefore, in the gauge $A^{a0}= \partial_i A^{ai}=0$, the
generating functional for the Schwarz--Sen approach is given by
\begin{equation}\label{36}
\tilde Z= \int {\cal D} A_{i}^{a}\, {\cal D}\pi^{a}_{i} \,\delta
(\partial^i A^{a}_{i})\,\delta(\partial^i\pi^{a}_{i}) \det(\nabla^2) 
\delta (\pi^{ia}_{T} +
\frac{1}{2} \epsilon^{ab} \epsilon^{ijk}\partial_j A^{b}_{kT}) 
{\det}^{1/2}(\epsilon^{ab} \epsilon^{ijk}\partial_k){\rm e}^{i\tilde S_eff}
\end{equation}
where
\begin{equation}\label{36a}
\tilde S_{eff} = \int d^4 x( \pi_{ia} A^{ia} - \frac{1}{2}(\nabla \times 
\vec A^{a})^2)
\end{equation}
\noindent 
and the trivial sector $A_{0}^{a}, \, \pi^{a}_{0}$ has already been
integrated out. We see that the $\det(\nabla^2)$ factor in (\ref{36})
arises also in (\ref{33}) due to the ghost contribution. Moreover as
\begin{equation}\label{37}
\det(\epsilon_{ab} \epsilon_{ijk}\partial^k) = 
{\det}^2(\epsilon_{ijk}\partial^k),
\end {equation}
we see, from (\ref{27}) that the two generating functionals $Z$ and $\tilde Z$
are actually identical.

We have thus shown the quantum equivalence of Schwarz--Sen and Maxwell
theories. This was expected as  both are free field theories that are
classically equivalent. 
Notice that the field equations were not used. Notice also that,
although the off-shell  descriptions look different, the physical content
of both formulations are the same since their generating functionals are
equal.
The important point is that the Schwarz--Sen approach can be understood as a 
way to implement duality as a local symmetry in Maxwell theory. This is 
possible thanks to the
introduction of the auxiliary fields $C_i$ and $P_i$. There are, of course,
other possibilities to turn the duality transformation local and presumably 
this may be related to the already known formulations where duality is 
realized locally.  We expect that this method may be also used for the 
interacting case since in essence it replaces nonlocal terms by auxiliary 
fields in a BRS invariant way.

\end{document}